\documentclass[fleqn,10pt]{wlscirep}

\usepackage{todonotes}
\usepackage{afterpage}
\usepackage{pdflscape}
\usepackage{xcolor}

\newcommand{\eg}{\emph{e.g.}\,}
\newcommand{\ie}{\emph{i.e.}\,}

\newcommand{\ket}[1]{\left|#1\right\rangle }

\newcommand{\braOket}[3]{\left\langle #1\middle|#2\middle|#3\right\rangle }

\newcommand{\onlinecite}[1]{\hspace{-1 ex} \nocite{#1}\citenum{#1}} 

\title{An investigation of pre-crystalline order, ruling out Pauli crystals and introducing Pauli anti-crystals}

\author[1,2,*]{Mikael Fremling}
\author[2,3]{J.K.~Slingerland}
\affil[1]{Institute for Theoretical Physics, Center for Extreme Matter and Emergent Phenomena,
Utrecht University, Princetonplein 5, 3584 CC Utrecht, the Netherlands}
\affil[2]{Department of Theoretical Physics, Maynooth University, Maynooth, co.~Kildare, Ireland}
\affil[3]{Dublin Institute for Advanced Studies, School of Theoretical Physics, 10 Burlington Rd, Dublin, Ireland}

\affil[*]{micke.fremling@gmail.com}

\affil[+]{these authors contributed equally to this work}

\begin{abstract}  
Fluid states of matter can locally exhibit characteristics of the onset of crystalline order.
Traditionally this has been theoretically investigated using multipoint correlation functions. 
However new measurement techniques now allow multiparticle configurations of cold atomic systems to be observed directly.
This has led to a search for new techniques to characterize the configurations that are likely to be observed.
One of these techniques is the configuration density (CD), which has been used to argue for the formation of "Pauli crystals" by non-interacting electrons in \eg a harmonic trap. 
We show here that such Pauli crystals do not exist, but that other other interesting spatial structures can occur in the form of an "anti-Crystal", where the fermions preferentially avoid a lattice of positions surrounding any given fermion.
Further, we show that configuration densities must be treated with great care as naive application can lead to the identification of crystalline structures which are artifacts of the method and of no physical significance.
We analyze the failure of the CD and suggest methods that might be more suitable for characterizing multiparticle correlations which may signal the onset of crystalline order. 
In particular, we introduce neighbour counting statistics (NCS),
which is the full counting statistics of the particle number in a neighborhood of a given particle.
We test this on two dimensional systems with emerging triangular and square crystal structures.
\end{abstract}
\begin{document}

\flushbottom
\maketitle
\thispagestyle{empty}

\section*{Introduction}
In characterizing the phases of multi-particle systems,
an important approach has long been to consider the real-space configurations of particles which contribute to the system's quantum ground state, or thermal state.
Often this proceeds through Monte Carlo (MC) averaging to a calculation of correlation functions,
particularly the (single particle) density and the two particle correlation functions,
as these can be rather directly related to conventional experiments.
In cold atomic systems on optical lattices it is possible to directly measure actual multi-particle configurations of bosons\cite{bakr2009quantum} or fermions\cite{Parsons15},
by measuring the positions of all particles in a single shot.
This creates an interest in characterizing more fully the configurations that might appear in such experiments.
This would be of particular interest in small, low dimensional or low temperature fluid systems,
where thermal and quantum fluctuations can easily wash out any local structure in the average density, but perhaps not in multi-particle correlations. 
Of course one may use multi-particle correlation functions for this,
but these have many parameters (the positions of multiple particles), which makes them harder to investigate and interpret.
We will here consider several alternative ways to investigate local order. 

In particular, we consider the \emph{Configuration Density} (CD) proposed by Gajda and collaborators\cite{Gajda2016EPL}.
This was used in a number of papers to suggest that configurations observed in fermionic systems, even without interactions,
should show  crystal like features, dubbed \emph{Pauli crystals}\cite{Gajda2016EPL,Rakshit2017} (see also  Ref. \onlinecite{Batle2017}).
The CD was furthermore used to support the idea that  configurations featuring in the wave functions of fractional quantum Hall (FQH) systems (famously quantum liquids) could show a high degree of crystalline order\cite{Lydzba2018AnnPhys}. 
While quantum Hall systems in condensed matter do not allow for direct observation of the individual electrons' positions,
analogues of integer quantum Hall states (Chern insulators) have already been created in atomic systems on optical lattices\cite{Aidelsburger2013,Miyake13},
which suggests that such observations may be possible in the near future. 

We will show first of all that particle configurations which might be considered Pauli crystals will most likely never be observed in the systems of interest.
Then we will investigate the CD's effectiveness as a tool for detection of incipient crystalline order.
We find that it can be very misleading,
showing clear ordered structure which is actually an artifact of the method.
In fact it was already noted in Ref.~\onlinecite{Gajda2016EPL} (figure 3) that the method can exhibit a bias towards a chosen seed configuration and we present a more systematic investigation of this bias here.

We then consider several other methods to investigate pre-crystalline order, which are more robust and less prone to artifacts.
The most interesting of these is the full counting statistics of the particle number in a disk of varying radius,
given that the disk contains a particle at its center.
We call this \emph{Neighbor Counting Statistics} (NCS).
We show that the NCS gives interesting information on the local neighborhoods of the particles,
which is complementary to the information obtained from the locations of peaks in the two-particle density.
It allows \eg for a clear identification of shells of neighbors.

Full counting statistics is currently becoming a popular way of characterizing the phases of multi-particle systems. In particular, in the context of cold fermionic gases,
there has been theoretical study of the full counting statistics of the particle number in spatial regions of the system,
see \eg Refs.~\onlinecite{Belzig2007,Humeniuk17} as well as actual measurement of the counting statistics\cite{Mazurenko2017}.
However, it appears that NCS have not been considered before.

\subsection*{Multi-particle systems considered}
In this paper we consider four types of multi-particle systems. 
The first system is a \emph{control} system of particles which are independently placed on a disk of radius $R$ according to the uniform distribution.
This will be useful for detecting artifacts in any of the methods we employ.

The second is the ground state of a system of non-interacting fermions in a two-dimensional harmonic trap.
This is also the system which features in Refs.\,\onlinecite{Gajda2016EPL,Rakshit2017}, where Pauli crystals are studied.  
To be concrete, the Hamiltonian of the system is $H=\sum_{j=1}^{N} \left( \frac{1}{2m}|\mathbf{p}_j|^2 + \frac12 m\omega |\vec{r_{i}}|^2\right)$. The  single particle eigenstates are given by
\begin{equation}
\psi_{n_x,n_y}\propto e^{-\frac{1}{2\xi^2}(x^2+y^2)}
H_{n_x}\left(\frac x\xi \right)
H_{n_y}\left(\frac y\xi\right),\label{eq:HO}
\end{equation}
where $\xi=\sqrt{\frac\hbar{m\omega}}$, $H_n$ is a Hermite polynomial, and $n_x,n_y\geq 0$ are integers.
The single particle energy is $\epsilon_{n_x,n_y}=\hbar\omega(n_x+n_y+1)$.
The $N$-particle state is constructed by filling a Slater determinant with the orbitals that minimize the total energy.

The third system is a two dimensional electron system.
Electrons in two dimensions under the influence of a strong magnetic field form integer\cite{Klitzing1980} and fractional\cite{Tsui1982} quantum Hall states,
which are gapped quantum liquids with quantized Hall conductivity.
At even higher magnetic field strengths, the electrons form solid phases,
or Wigner crystals (Wigner already appreciated this possibility in 1934\cite{Wigner1934}).
There is strong theoretical (see \eg Refs. ~\onlinecite{Lam1984,Zhao18PRL}) and experimental\cite{Andrei1988,Goldman1990,Jiang1991,Williams1991,Li1991,Buhmann1991,Paalanen1992,Pan2002,Maryenko18} evidence that this happens at fields and electron densities corresponding to Landau level filling factors of roughly $\frac{1}{5}$ and below.
At lower filling fractions, one finds solid and fractional quantum Hall liquid phases alternating downward from there to about filling $\frac{1}{9}$.
The precise transition to a Wigner Crystal depends in a complicated way on \eg sample quality and temperature,
where the disorder potential plays an important role by pinning the crystal.
There are also glassy and localized phases and ``intermediate" phases between crystal and liquid.
One may consult Refs.~\onlinecite{Fertig1997,Shayegan1997} for early reviews of this rich subject.

Rather than consider all the complications of the real system we can instead consider the system described by the Laughlin wave function\cite{Laughlin1983} 
\begin{equation}
\Psi_{L}^{(q)}=\prod_{i<j}(z_i-z_j)^{q}e^{-\sum_{i}|z_i|^2/4\ell^{2}}.\label{eq:Laughlin}
\end{equation}
Here the $z_{j}=x_{j}+i y_{j}$ are complex coordinates in the plane, $q$ is a parameter,
which fixes the filling fraction $\nu$ at $\nu=\frac{1}{q}$ and $\ell$ is the magnetic length. 
For $q=1$, this is the ground state of the non-interacting electron system at $\nu=1$ (it is just the usual Slater determinant in a different guise) and hence a good test system to search for Pauli crystals. For $q>1$ odd,
$\Psi_{L}^{(q)}$ is the ground state of an electron system with short range interactions, with the electrons confined to the lowest Landau level.
It is also an excellent trial wave function for the ground states of the fractional quantum Hall liquids at fillings $\nu=\frac{1}{3}$ and $\nu=\frac{1}{5}$,
with $q=3$ and $q=5$ resp.
For $q>7$, $\Psi_{L}^{(q)}$ is no longer a good trial wave function for the system with Coulomb interactions,
but nevertheless a good test system for us.
The probability density $|\Psi_{L}^{(q)}|^2$ is identical to the probability density for a \emph{classical} two dimensional Coulomb plasma (with logarithmic interaction potential). This has been studied in great detail 
\cite{Caillol1982,DeLeeuw1982} and it shows a phase transition from a liquid to a solid phase at $q\approx 70$,
so we can study liquid and solid behaviors within this family of states. 

Finally, we consider a classical system of $N$ particles on the plane,
interacting through a modified Lennard-Jones potential, 
\begin{equation}
V(r)=\frac{1}{r^{12}}-\frac{2}{r^6}-\epsilon \exp\left(-\frac{(r-r_0)^2}{2\sigma^2}\right).\label{eq:LJ-pot}
\end{equation}
The first two terms form the usual Lennard-Jones potential,
while the term proportional to $\epsilon$ introduces a second minimum near $r=r_0$,
whose depth and width can be adjusted using $\epsilon$ and $\sigma$. 
This potential was introduced by Engel and Trebin\cite{Engel2007} to study complex crystalline and quasicrystalline phases in two dimensions using only a single type of particle.
They found a complicated phase diagram with many different solid phases exhibiting different crystal and quasicrystal structures at low temperatures.
Naively, the ratio of distances between the first and second minimum of the potential should correspond to the ratio of nearest neighbor to next to nearest neighbor distances on the solid lattice.
In this paper,
we will only consider the parameter values $\epsilon=1.1$, $\sigma^2=0.02$ and $r_0=1.4$.
For these values the system forms a square lattice at low temperatures.
We will consider temperatures near the melting point to study the local emergence of the square lattice.

\begin{figure}[ht]
  \centering
  \includegraphics[width=1.00\linewidth]{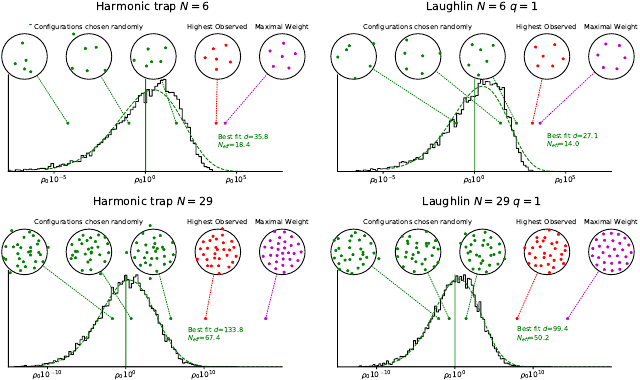}
  \caption{Actual configurations for $N=6$ (upper) and $N=29$ (lower) electrons of the Harmonic trap (left) and Laughlin $\nu=1$ state (right).
  The panels show the \emph{maximal weight} configurations (rightmost),
  \emph{maximal weight} configurations actually obtained in the MC simulation of the system (second from right)
  and three randomly chosen configurations (left).
  These are marked in the histogram of the corresponding probability density distribution.
  $\rho_0$ is the geometric average of the probability densities of the generated MC samples.
  In all panels $10^4$ independent configurations where generated.
  For comparison, we also plot the function $\rho\ln(1/\rho)^{\frac{d}{2}-1}$,
where $\rho$ is the probability density (see main text for a justification).
  Note how randomly chosen configurations show very little signs of forming a crystal.
  Even the state with the highest observed probability density, among the $10^4$ MC samples,
  is quite far from the \emph{maximal weight} configuration. 
}
\label{fig:confs}
\end{figure}

\section*{Results}

\subsection*{Against Pauli crystals}

We start off with an argument against the physical relevance of Pauli crystals. 
In Refs.\,\onlinecite{Gajda2016EPL,Rakshit2017} the authors proposed that  multi-particle states in certain systems of non-interacting fermions should show crystal-like patterns, which could be seen in experiments,
due to the effective inter-particle repulsion caused by the Pauli exclusion principle.
We now argue that such "Pauli crystals" are actually unlikely to ever be observed.
In fact, the typical configurations that would be observed,
for both the $\nu=1$ Laughlin state and the 2D-harmonic trap,
show no discernible crystalline ordering.
Clearly, due to the fermion exclusion, there is a real-space correlation hole around each particle, but we now show
that this does not lead to a crystal-like order.

Our point here is that there is a low \emph{probability} to observe configurations with a very high \emph{probability density}.
  This is caused by the fact that there are so many more configurations with a lower probability density.
  It is a similar situation as in classical systems at nonzero temperature,
  where the lowest energy state always has the highest Boltzmann probability density, but one typically observes higher energy configurations as these have higher entropy.

To make this clear we first produce a large number of configurations from the wave functions (one of which is considered in Ref.~\onlinecite{Gajda2016EPL}) using importance sampling (by Metropolis Monte Carlo methods).
It is crucial to realize that the sample configurations are produced according to the quantum mechanical probability distribution given by the probability density $|\Psi(x_1,...x_N)|^2$, which is the same probability density $|\Psi|^{2}$ that would govern the experiment.
In other words, the configurations we observe in the simulation are representative for the configurations that would be observed in an experiment.
  
Once we have the samples, we can plot a histogram of their probability densities.
This is shown in Fig. \ref{fig:confs} for $N=6$ particles (upper panels) and for $N=29$ particles (lower panels).
Note that the probability density (on the horizontal axis) is on a logarithmic scale.
Here the vertical axis measures how many configurations exist with a given probability density, \ie it measures the \emph{probability density} in \emph{probability density space}.
I other words, we make bins on the probability density axis and plot the number of configurations in our simulation which have probability density that falls within each bin.
The bins are of equal size when viewed on the logarithmic scale in the plot.
For large numbers of bins and samples, the resulting plot approaches (up to global normalization) the probability density on the one dimensional space which is parametrized by the logarithm of the probability density itself.
We see that the observed probability densities are sharply peaked in probability density space, but not at the maximal probability density. 
We also observe that no configurations are found with probability density near that of the highest probability density configuration
(purple configuration, furthest right, obtained using the Nelder-Mead method\cite{Nelder1965}),
indicating that configurations with such high probability density actually have negligible probability of being observed.

The probability density is a good indicator of how well separated the particles are and therefore a low probability density configuration typically does not show any clear crystalline structure.
Indeed, direct observation of randomly chosen samples (green configurations, left) shows no configurations with anything resembling crystalline ordering of the particles.

The MC sample with the highest probability density that was actually found among the MC samples
(red configuration, second from right) still has a value of $|\Psi|^2$ which is orders of magnitude smaller than the \emph{maximal weight configuration}.
These are somewhat more regular than the typical configuration, but still much less regular than the configuration with the highest probability density.

Of course only two system sizes are shown here as an illustration, but results at other sizes are similar.
For really small systems, which where the focus of Ref.~\onlinecite{Gajda2016EPL}, the likelihood of regular configurations is increased already by virtue of the smaller configuration space alone. 
For instance, for a system of only three particles, it is nearly inevitable that there will be some probability to observe configurations where the particles are approximately located on the vertices of an equilateral triangle.
It is also likely that this probability may be somewhat enhanced in fermionic systems,
as the wave function necessarily has zeros when the particle positions coincide.
Nevertheless we have run our simulations also for three particle systems and we find that, while there is now a nonzero probability of finding configurations with high probability density,
where the particles are approximately on an equilateral triangle,
this probability is still relatively small, even in systems of fermions,
and less ``crystalline'' configurations occur more frequently.

One may wonder if there is a qualitative difference if one considers the sequence of ``magic numbers'' $N=1,3,6,10,\ldots$ for which the harmonic oscillator potential affords a unique ground state.
This however, makes little difference.
We have data also for the ``magic number'', $N=28$ and this data is qualitatively very similar to the $N=29$ and $N=30$ cases shown in the figures.

To analytically understand the suppression of the \emph{maximal weight} configurations, we introduce a simple heuristics model of the probability density.
The idea is to assume that the  probability density varies as a $d$-dimensional Gaussian around the \emph{maximal weight} configuration,
where $d$ is an effective dimension which we use as a fit parameter.
Clearly this is a very crude model as the actual parameters of the model (the positions of the particles) will have complicated correlations.
Nevertheless, any probability density with a global maximum can be at least locally approximated in this way near the maximum.
For the assumed distribution, we can calculate the histogram (see the Methods section for the details).
We then expect the histogram in our figure (the probability density in probability density space) to be described by the function $\rho\cdot \ln(1/\rho)^{\frac d2-1}$, where $\rho$ is the probability density.
By fitting against the actual histogram we can estimate the number of degrees of freedom $d$,
and obtain the green lines in Figure~\ref{fig:confs}.
For $N$ particles in two dimensions in a rotationally symmetric state we would naively expect $d=2N-1$ and we note that this estimate is roughly a factor of two away from the best fit in our plots. 
Defining $N_{eff}=\frac{d+1}{2}$ we see, at $N=29$ for example,
that $N_{eff}\approx67.4$ and $N_{eff}\approx50.2$ for the Harmonic oscillator and $\nu=1$ state respectively.
We should stress once more that also in this crude model, where we assume a single configuration with maximal probability density and Gaussian decay of the probability density away from that configuration,
we find that the actual configurations found when drawing from this probability density typically have much lower than maximal probability density. In fact, we show in the Methods section that, in this model,
the distance between the peak of the histogram and the maximum of the probability density is proportional to $N_{eff}$,
while the width of the peak only scales as $\sqrt{N_{eff}}$, both on the logarithmic scale of the plots.
This strongly suggests that very high probability density configurations become increasingly unlikely to be observed for larger systems.      

\subsection*{The Pauli Anti-Crystal}
The conclusion that Pauli crystalline configurations would be practically impossible to observe should really not come as a surprise.
In fact, for free fermions we now present a simple general argument against Pauli crystals.
The impossibility of Pauli crystals does however not prevent interesting correlations,
and one may for instance hope to find a kind of Pauli anti-crystal,
where the Pauli principle prevents two electrons from being in certain positions relative to each other.
The correlation hole around each electron is a simple example of this, but as we shall see below, more elaborate anti-correlation behaviour is possible.
For this argument it is important to clarify what we require of a Pauli crystal. 
First of all, it is a system of \emph{non-interacting} fermions, that is, the only interaction is the statistical interaction implemented by the anti-symmetry of the wavefunction. We implement this condition by requiring that the system is in a Fock state, or equivalently, its wave function is described by a single Slater determinant. Obviously this may exclude some degenerate cases.  Secondly, the system should show spatial features reminiscent of a crystal in configurations that would be observed in experiment. This should be reflected in multipoint configuration functions. Since the multipoint functions for a single Slater determinant are all determined from the the density (one-particle correlation function) and the two-particle correlation functions (Wick's theorem\cite{wick1950}), we can focus on these. Crucially we require that the single particle density does not show the crystal structure. This density can be written as the sum of the spatial probability densities of the occupied orbitals. Hence if there was crystal structure in the density, it would mean that the selected orbitals are already restricted to be supported on a crystal. In such a case, the crystal structure has nothing to do with the statistical interaction, but is purely due to the single particle potential. In particular, one would see the crystal even if the particles were distinguishable, or bosons.
Therefore we can think of a Pauli crystal as a system of non-interacting fermions, described by a Slater determinant wavefunction, that shows no spatial features in the density, but has crystal like signatures in the two-particle density.

To make this more concrete, consider the Fock state
$\ket{\Psi}=\prod_{k\in K}a_{k}^{\dagger}\ket{\Omega}$, where $\ket{\Omega}$ is the vacuum state, and $K$ is the set of orbitals that are filled.
We define the creation operator in position space as
$\hat{\psi}^{\dagger}(\vec x)=\sum_{k}a_{k}^{\dagger}\psi_{k}(\vec x)$
and the density operator as
$\hat{\rho}(\vec x)=\hat{\psi}^{\dagger}(\vec x)\hat{\psi}(\vec x)$,
where $\psi_k(\vec x)$ is the orbital wave function labeled by $k$.
The one-particle density is in this case given by $\rho(\vec x)=\braOket{\Psi}{\hat{\rho}(\vec x)}{\Psi}=\sum_{k\in K}|\psi_{k}(\vec x)|^2$.
The two-particle correlation function can be written as  
\begin{equation}
\rho(\vec x,\vec y) = \braOket{\Psi}{\hat{\rho}(\vec x)\hat{\rho}(\vec y)}{\Psi} =\rho(\vec x)\rho(\vec y)-\left|\rho(\vec x \to \vec y)\right|^{2}+\delta(\vec x -\vec y)\rho(\vec x),\label{eq:rho_x_y}
\end{equation}
where 
\begin{equation}
\rho(\vec x \to \vec y)=\rho(\vec y \to \vec x)^{\star}=
\sum_{k\in K}\psi_{k}\left(x\right)\psi_{k}^{\star}\left(y\right).
\label{eq:rho_x_to_y}
\end{equation}
The last term in \eqref{eq:rho_x_y} is the auto-correlation term, which accounts for the fact that a
particle at $x$ is its own neighbor, by definition.
An important observation here is that if we integrate the three terms over all $\vec x$ and $\vec y$ they evaluate to $N^2-N+N=N^2$.
In particular, $\left|\rho(\vec x \to \vec y)\right|^{2}$ integrates to $N$. 
It is clear that $\rho(\vec x,\vec y)$ is positive and clearly so is $\left|\rho(\vec x \to \vec y)\right|^{2}$.
Hence we have $0\leq \rho(\vec x,\vec y) \leq \rho(\vec x)\rho(\vec y)$ (for $\vec x\neq \vec y$).
In other words, the two particle correction function at $(\vec x,\vec y)$ can never exceed the product of the density at $\vec x$ and at $\vec y$.

In a Pauli crystal we require that $\rho(\vec x)$ has no spatial structure, \ie $\rho(\vec x)=\rho_0$, at least in the bulk of the system, to good approximation.
We may construct the usual correlation function $g(\vec r)$ as 
\[
g(\vec r) = \frac1N \int dR \rho(R+r,R) = \rho_0 - f(\vec r) + \delta(\vec r), 
\]
where $f(\vec r)=\frac{1}{N}\int dR\,\left(\sum_{k,k^{\prime}\in K}\psi_{k}(\vec R+\vec r)\psi_{k^{\prime}}^{\star}(\vec R+\vec r)\psi_{k^{\prime}}(\vec R)\psi_{k}^{\star}(\vec R)\right)$.
This means that all the nontrivial behaviour is in $f$,
which comes from the integral of the second term $\left|\rho(\vec x \to \vec y)\right|^{2}$.
Since $\int dr f(\vec r)=1$, it cuts exactly one particle out of the total of $N$ which would be given by the constant density $\rho_0$.
Normally we would expect this to happen near $r=0$, with a maximum in $f$ ``digging out" a correlation hole there.
If we want to see obvious crystalline features in $g$, we would necessarily have to evacuate particles (mostly) from some large region, away from the lattice positions, or away from shells that are obtained after rotational averaging of the lattice.
Since $f$ can only remove a total of one particle and there is no other mechanism to get away from the constant density $\rho_0$, this is not possible.  
Nevertheless, we can have interesting features in $\rho$ as correlation holes will appear whenever $f(\vec r)$ has a maximum.
Such a maximum should always appear at $\vec r=0$, but there can be more, depending on the chosen set of $\psi_k$.
In such a situation, the usual single correlation hole is replaced by several smaller disjoint correlation holes, still together accounting for a total deficit of one particle.
It is possible for these correlation holes to be arranged in a crystalline pattern, and we call this a \emph{Pauli Anti-Crystal}.
It would be an interesting challenge for experiments to engineer these anti-crystals in \eg cold atom experiments.

\begin{figure}[t]
  \centering
  \includegraphics[width=.90\linewidth]{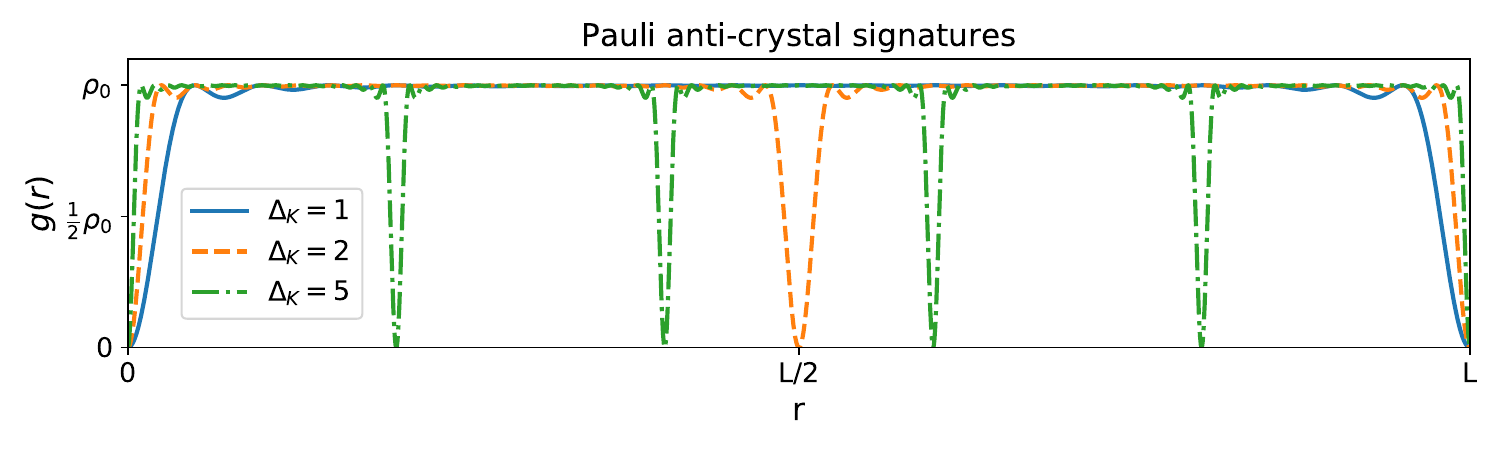}
  \caption{Examples of extra correlation holes in a 1D Pauli Anti-Crystal with $N=20$ particles.
    By constructing Slater determinant of plane waves $\psi_{k}(x)=\frac{1}{\sqrt{L}}e^{i 2\pi k\frac{x}{L}}$ with $k=\Delta_k n$, where $n=1,\ldots,N$ dips in $g(x,y)$ are observed when $x-y$ are multiples of $\frac{L}{\Delta_k}$.
    Notice how the integrated volume of $\rho(x,y)/N$ is exactly $(N-1)$ (if we exclude the delta function distribution at  $x=y$) which means that the correlation hole accounts for precisely $1$ particle.
    The correlation hole volume can thus not be made larger or smaller, but it can be reshaped, or even spit into several pieces.
    }\label{fig:AntiPauli}
\end{figure}

As a concrete example of a Pauli anti-crystal consider for simplicity a one-dimensional system on a circle of length $L$.
We now choose the set of $\psi_{k}$ such that $\rho(x)$
is completely flat, \ie $\rho(x)=\rho_{0}$.
This is accomplished easily by choosing the set of plane wave functions $\psi_{k}(x)=\frac{1}{\sqrt{L}}e^{i 2\pi k\frac{x}{L}}$, which gives $\rho(x)=\rho_0=\frac{N_{e}}{L}$.
However, $\rho(x \to y)$ will be quite different since
\begin{align*}
\left|\rho(x \to y)\right|^{2} = 
\frac{1}{L^{2}}\left|\sum_{k\in K}\cos\left(2\pi k\frac{x-y}{L}\right)\right|^{2}+
\frac{1}{L^{2}}\left|\sum_{k\in K}\sin\left(2\pi k\frac{x-y}{L}\right)\right|^{2}.
\end{align*}
 By choosing the values of $k$ to be $k=\Delta_k n$, where $n=1,\ldots,N$ we can create perfect constructive interference when $x-y$ is a multiple of $\frac{L}{\Delta_k}$.
 See Figure.~ \ref{fig:AntiPauli}, where we have plotted $g(r)=\rho(R\to R+r)\frac{L}{N}$ for the three cases $\Delta_k=1,2,5$ for $N=20$ particles. We can clearly observe there that the size of the disjoint correlation holes decreases as their number increases.

 The reader might wonder if there is a relation between the material in this
 section and the Javanainen method\cite{Javanainen1996} discussed in Ref.~{\onlinecite{Gajda2016EPL}}, but that is not the case.
 The Javanainen method studies a given wave function by fixing successive particles in the positions which maximize the conditional real space probability density, given the positions of the particles that were fixed earlier.
 Our plots of the anti-crystal states are simply standard two-particle correlation function plots and we do not fix the position of any particle to generate these.

\subsection*{The Configuration Density}

One of the tools that was employed to support the idea of Pauli crystals is the configuration density (CD).
This is an intuitively attractive tool for detecting local crystalline order,
in a system that does not show this order directly through fluctuations in the actual density.
 However, we will now show that it can be highly misleading. 

To obtain the configuration density (CD) of a multi-particle system,
one needs a set of representative configurations (sets of positions of all particles),
for example obtained by MC sampling of the system's wave function
or thermal distribution, or even by direct experimental observation.
The basic idea for the CD is that these representative configurations of the system exhibit spatial order,
but symmetry transformations (rotations and translations) wash it out in the average.
To prevent this from happening, one chooses a suitable seed configuration (SC).
Usually this is taken to be the configuration with the largest probability density, which often does exhibit a regular spatial structure.
Then one defines a quantity which characterizes how close configurations are to each other.
We might call this the configuration distance, see formula \eqref{eq:confdist} below. 
All representative configurations are now rigidly rotated and shifted to minimize their configuration distance.
The configuration density is then calculated by taking the usual density from the resulting rotated and shifted configurations.

\begin{figure}[t]
  \centering
  \includegraphics[width=1.00\linewidth]{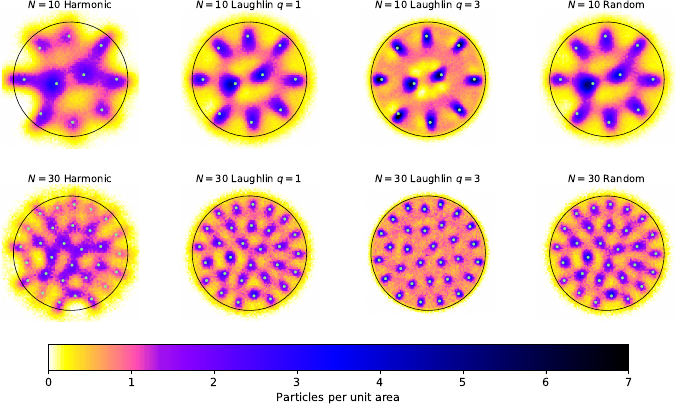}
  \caption{Configuration densities for (from left) 2D-Harmonic trap,
Laughlin $q=1$, Laughlin $q=3$ and uniformly distributed random points.
    The upper row is for $N=10$ and the lower is for $N=30$.
    The number of MC points is $10^5$ and $10^4$ respectively.
    All CD:s (of the same size) use the same seed configurations.
    It is clearly visible how the pattern of the seed configuration is imprinted on all DC:s, even on the random points.
    The color scale on all plots is chosen such that area of the circles are equal to the number of particles,
\ie there is on average 1 particle per unit area in all plots.
    }
  \label{fig:CDs}
\end{figure}

\begin{figure}[t]
  \centering
  \includegraphics[width=1.00\linewidth]{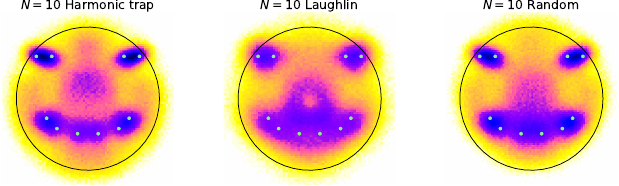}
  \caption{The CD with a smiley as a seed for $N=10$ particles and $10^5$ Monte Carlo samples.
      The states shown (from the left) are 2D-Harmonic trap, Laughlin $q=1$ and uniformly distributed random points.
    The color scale and length scale is the same as in Fig.~\ref{fig:CDs}.
  }
\label{fig:smile_face}
\end{figure}

Often the CD will have a remarkable amount of clearly visible structure,
even if the usual density, calculated from the unprocessed configurations, does not. 
Most notably one observes peaks about the locations of the particles in the SC.
It is tempting to take this as an indication that the configurations of the system generally show a crystalline order similar to that of the SC, just shifted and rotated. However,
we will show that peaks around the SC remain present even if the particle positions in the sample configurations are uncorrelated and chosen according to a uniform distribution, see Fig. \ref{fig:CDs}.
Moreover, passing to systems with interactions and/or fermionic statistics,
we find that one may choose different seed configurations in obviously unlikely or physically implausible ways, for example in the shape of a smiley face,
and prominent peaks in the CD will still appear near the seed positions, see Fig. \ref{fig:smile_face}.
All this suggests that visual identification of structure in the CD is not a good way to characterize the configurations relevant to actual physical systems.

In the rest of this section we make an effort to analyze the CD in more detail and to supplement it with a new measure of particle correlations, the \emph{configuration variance} (CV),
in order to see if ideas along the lines of the CD may in fact yield more reliable physical information.

\subsubsection*{Distance between configurations}
An essential ingredient for the method of configuration densities is a measure of distance between two configurations.
In Ref.~{\onlinecite{Gajda2016EPL}} the measure of closeness was defined using only the angular coordinates of the particles and required an identification of the particles in the two configurations,
inducing a permutation on the coordinates.
This method is only feasible for small numbers particles and in cases where there is a clear shell structure in the positions.
We therefore consider an alternative measure introduced in Ref.~{\onlinecite{Lydzba2018AnnPhys}}.
For two configurations $\vec{r},\vec{r}'$,
the configuration distance $g(\vec r | \vec r^\prime)$ is defined by
\begin{equation}
\label{eq:confdist}
g(\vec r | \vec r^\prime) = \sum_{i,j=1}^N e^{-\frac1{G^2}(r_i-r^\prime_j)^2}. 
\end{equation}
This can be easily calculated, with scaling in time of order $\mathcal O(N^2)$, and is bounded in the interval $0<g\leq N^2$.
The heuristic is simply that the closer two configurations are to to each other, the more the Gaussians will contribute.
The measure has a parameter $G$, which controls the length scale over which two particle positions are considered similar.
We find that it is best to take $G$ comparable to the inter-particle separation. If $G$ is much larger, then
$g(\vec r | \vec r^\prime)$ becomes insensitive to the local structure of the configurations,
and if $G$ is much smaller than this,
the final position of the optimized configuration often just depends 
on the individual particles which happen to be very close to a seed position after rotation.

To generate the data used in the plots of the configuration density shown in this paper,
we have optimized each configuration $\vec{r}$ only by rotating it to minimize the configuration distance $g(\vec{r}|\vec{r}')$ to the sample configuration $\vec{r}'$.
We did not apply any shift to the sample configurations.
In the work of Gajda\cite{Gajda2016EPL}, all configurations were shifted to make their centres of mass coincide with that of the seed configuration.
While this does not introduce a new fitting parameter it does likely improve the fit between the samples and seed a little as at least the centres of mass are made to match.
However, in other works, it appears that shifts are not always treated in the same way,
for example, the paper by Lydzba and Jacak\cite{Lydzba2018AnnPhys} does not appear to apply shifts at all.
In any event, there is limited scope for shifting to make much of a difference as the centre of mass only varies very slightly between configurations (except at very small numbers of particles).
To make sure of this, we have also implemented the shift which makes all centres of mass coincide,
and confirmed that it makes no qualitative difference (and usually no discernible difference) to the configuration density plots shown here.

We also take as convention that we rescale the seed configuration $\vec r^\prime$ such that $r^2$,
averaged over the SC is the same as $r^2$ averaged over the full set of Monte Carlo configurations.
This allows us to use the same seed configuration for different systems.
A ring of radius $\sqrt{2\left<r^2\right>}$ is drawn to simplify comparisons of the plots.

\subsection*{Configuration Densities are highly biased by the seed configuration}
We now demonstrate that the Configuration Density (CD) is highly biased by the seed configuration (SC).
We do this in two different ways:
First we show that even random points will create a CD that mimics the SC.
This can be seen in Figure~\ref{fig:CDs}.
In this figure, we have constructed the configurations densities for the 2D-harmonic trap,
the $\nu=1$ and $\nu=1/3$ Laughlin states and for independent uniformly distributed particles,
for both $N=10$ and $N=30$ particles.
We use the same seed configurations in all cases,
namely the maximum likelihood configuration for the family of Laughlin states.
This configuration is actually independent of $q$.
In the figure we can clearly see how the SC is imprinted on the CD for all the examples considered,
even the random state.
This directly shows that it is possible to find a CD that apparently shows crystalline order, even though it is not really there.

As a second example, we show that not only can we imprint a SC on a random state,
it is also possible to imprint physically unmotivated, or bizarre SCs onto a CD for a system of fermions, which might be expected to show physical structure.
To illustrate this point we show in Figure~\ref{fig:smile_face},
how a smiley face of $N=10$ particles can be imprinted not only on the random state,
but also on the Harmonic oscillator state and the Laughlin $q=1$ state.

From these tests we can conclude that naive use of the CD method can result in a strong bias towards the seed configuration.

\subsection*{Further analysis of the CD -- the CV}

We now discuss briefly what causes the misleading results of the previous section.
We also introduce a measure that can be used to complement the configuration density when searching for hidden orders, the \emph{configuration variance} (CV).

It is natural to ask why we see so much structure in the CD even when it is not reflected by the sample configurations.
We conjecture that this is essentially due to the relatively large density fluctuations in the samples (clusters of particles and voids in the configurations). During the rotation of the samples to create the CD,
clusters of multiple particles have high probability to be rotated near a seed position. Similarly,
voids will be preferentially located away from the seed positions. As a result,
the average configuration density near the seed points is enhanced, giving rise to the peaks in the CD. 

We can test this explanation by  considering the variances of the numbers of particles in neighborhoods of the seed configurations. 
The configuration variance $C_V(r)$ for a particular seed position is defined as the variance of the particle number in a disk of radius $r$ around that seed position,
calculated from the sample configurations, after the processing (optimization of configuration distance) which features in the calculation of the CD has been applied to them.
In effect,
\[ 
C_V(r)= \left< n^2 \right>_r -  \left< n \right>_r^2, 
\]
where $n$ is the particle number density and $\left< X \right>_r$ is the average of the quantity $X$ within a circle of radius $r$ around the seed configuration,
calculated from the processed sample configurations. Of course we will consider the CVs for all seed positions.

\begin{figure}[t]
  \centering
  \includegraphics[width=1.00\linewidth]{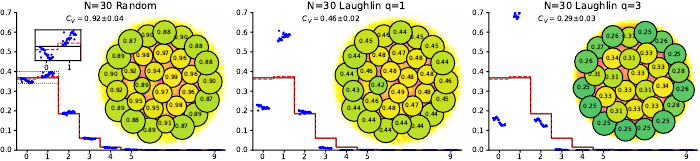}
  \caption{The variances of the numbers of particles present in circles centered on the seed positions and containing on average 1 particle.
Left: Random points, Middle: Laughlin state at $q=1$, Right: Laughlin state at $q=3$.
  Also plotted: The distributions of the numbers of particles in the neighborhoods of the seed points. 
  The distribution (histogram) of the number of particles around each of the seed points are marked with separate (blue) markers.
  For ease of viewing, the points are ordered from left to right according to increasing radial position of the seed points.
  The solid (black) line shows the Poisson distribution $P(k)=\frac{e^{-\lambda}\lambda^k}{k!}$, with the average $\lambda=1$.
  The dashed (red) lines shows the binomial distribution $P(k)={\binom{N}{k}}p^k\left(1-p\right)^{N-k}$, with  $p=\frac{1}{N}$, ensuring the appropriate density.
  }
  \label{fig:CV-plot}
\end{figure}

We can choose the radius $r$ of the disk around each seed position so that the disk contains an average of $m$ particles, so $\left< n \right>_r=m$.
Again this choice is based on the processed configurations and hence this radius can be obtained directly from the CD.
It then also makes sense to define $C_{V,m}$ to be the average of the values of $C_V(r)$ over all seed positions, with $r$ chosen individually for each seed position such that $\left< n \right>_r=m$.
A particularly interesting case is to choose $r$ in such a way that the disks contain a single particle on average ($m=1$). We write $C_{V,1}$ simply as $C_V$.

If the true system has a rigid crystalline order, then $C_V$ should be small,
as there would be little chance to observe particles far away from their lattice positions, which should be close to the seed positions.
Therefore there would be little chance to find multiple particles, or no particles,
in the neighborhood of any particular seed position. However, if the CD does not reflect lattice structure,
but simply the fact that clusters of particles are preferentially rotated to be close to seed positions,
then we would expect to find relatively large values of $C_V$.

We illustrate the configuration variance in Figure~\ref{fig:CV-plot}.
We have there adjusted $r$ around each seed position such that $\left< n \right>_r=1$ precisely.
We then measure $C_V$ for three different types of states:
random positions (left), the Laughlin $q=1$ state of free fermions (middle), and Laughlin $q=3$ state of interacting fermions (right).
In the figure we have drawn the circles of radius $r$ corresponding to $\left< n \right>_r=1$ around the seed positions,
and the number within each circle is the $C_V(r)$ for that site.
We also give the value of $C_V$, which is just the average of the values in the circles.  

First we consider the case of randomly distributed points, see the left hand panel of Figure~\ref{fig:CV-plot}.
We find that the configuration variance is in the range $C_V(r)\approx0.96\pm0.02$ for the $N=30$ points considered.
This is close to what one would get by assuming that particles are Poisson distributed within the disk of radius $r$, with average $\lambda=\left<n\right>_r=1$, as that would produce a variance of $C_V=\lambda=1$.
We recall that the Poisson distribution describes the number of completely uncorrelated random events,
so finding this here would mean that the distribution of particles in the circle is modeled well by the assumption that the event of finding some particle within the circle is completely uncorrelated with the event of finding any other particle within the circle.
This is clearly not indicative of lattice structure but very compatible with the fact that the configurations (before CD processing) were generated precisely by placing the particles on the system in a uniformly random and uncorrelated way. 
In fact, being more precise, we may note that there is some correlation between the events of finding the various particles within any disk,
since we know that there are a fixed number  of particles in total, $N=30$.
The most appropriate distribution is thus a binomial distribution, with parameter $p=\frac{\left<n\right>_r}N=1/N$.
The variance of this distribution is $C_V=Np(1-p)\approx0.967$,
which conforms to observations even better than the Poisson value.
To confirm our suspicion on the distribution, we also calculated the full probability distribution for the number of particles in each circle.
This is also graphed and we see that the binomial distribution is perfectly reproduced.
It is even possible to -- just barely -- discern the difference between the Poisson distribution (black line) and the Binomial distribution (red line). 

We now turn our attention to the cases of non-interacting and interacting fermions in the Laughlin states at $q=1$ and $q=3$ respectively,
shown in the middle and right hand panels of Figure~\ref{fig:CV-plot}.
The first observation we make is that $C_V$ is considerably lower than the value for randomly placed particles,
for both of the Laughlin states considered, with the higher $q$ state having a lower $C_V$.
Also on the level of the probability distributions $P(k)$, we see that the two Laughlin states deviate substantially from the Poisson/Binomial distributions.
Especially the $P(1)$ probability is enhanced, while the $P(0)$ probability is suppressed.
The $P(k>2)$ probabilities are suppressed relatively even more.
All this is in accord with the intuition that in the fermionic states, and especially in the strongly interacting $q=3$ state,
the particles are correlated in such a way that they cluster less.
This does not necessarily mean that there is crystalline order (there is not, as we showed earlier),
but nevertheless these states are \emph{incompressible} liquids and the CV and corresponding distributions $P(k)$ are able to pick up on this,
while the CD on its own does not show any obvious features which distinguish these states from our control system based on randomly placed points (cf.~Figure~\ref{fig:CDs}).

\section*{Neighbor Counting Statistics}

In the previous sections, we have seen that the introduction of a seed configuration can be more of a distraction than a help in discovering local correlations in multiparticle systems.
The variance and especially the full counting statistics of the particle number in a region surrounding one of the seed configurations were more useful.
However, as the  seed configuration could still bias the results,
it seems natural to apply these tools without any dependence on a seed configuration.
Of course one may simply study the counting statistics of any spatial sub-region of the system. This has been done before,
even in the context of cold fermionic gases\cite{Belzig2007,Humeniuk17},
and the particle number distribution can even be experimentally observed there\cite{Mazurenko2017}.
We can think of the full counting statistics of a region as a tool which naturally extends the single particle density, by including information about multi-particle correlations,
without actual direct calculation of multi-particle correlation functions.
From the density we can obtain the average number of particles in any given region,
but the full counting statistics gives much more information,
for example all the moments of the distribution, including the variance. 

Here we want to instead improve on the two particle correlation function,
which is of course a standard tool to detect local correlations in systems with homogeneous density (e.g. correlation holes around particles).
Therefore we will study the \emph{neighbor counting statistics} (NCS) of our systems.
The neighbor counting statistics are simply the full counting statistics of the neighborhood of a particle in the system.
To be concrete, we define $p_n(r)$ to be the probability of finding $n$ particles within a distance $r$ from another particle. Again,
the two point correlation function $g(r)$ already allows one to find the average number of particles within distance $r$ of a given particle,
but the NCS will give full information on the distribution of the particle number, which includes information on multi-particle correlations.
In particular we can easily get the variance of the number of neighbors from it.

Calculation of the NCS is straightforward, assuming that one can easily generate a large number of representative sample configurations for the system.
One simply counts the number of particles within distance $r$ of every particle in all sample configurations,
while keeping a histogram of the number of neighbors found, so in effect,
one keeps a tally of how often one finds $0$ neighbors, $1$ neighbor,
$2$ neighbors etc. within a distance $r$. After normalization,
the values in the bins of this histogram are the best estimate of $p_n(r)$ from the samples. 

We now calculate the NCS for the Laughlin states, which are known to be liquid at low $q$,
but which crystallize into a triangular (Wigner) crystal at high $q$.
We also consider classical particles interacting through the modified Lennard-Jones potential given in \eqref{eq:LJ-pot},
with the parameters set as indicated ($\epsilon=1.1$, $\sigma^2=0.02$ and $r_0=1.4$),
so that the system forms a square lattice at low temperatures. For both systems,
we can efficiently generate representative configurations by Monte Carlo importance sampling.
We will see that the NCS is quite helpful in identifying the incipient crystalline orders in these systems.

\subsubsection*{Laughlin State}

\begin{figure}
  \centering
  \includegraphics[width=1.00\linewidth]{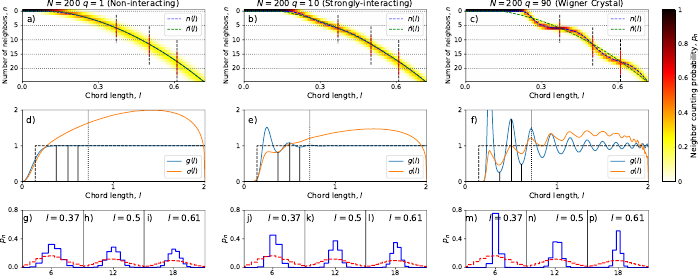}
\caption{Neighbor counting statistics for the Laughlin states, $N=200$, $q=1,10,90$.
  The blue line shows the average numbers of particles $n(l)$, within the chord length $l$.
  The green line is a guide to the eye, and shows and approximation of a sharp correlation hole around each particle.
  The precise form is $\tilde n(l)=Nl^2/4-1$ (for $l>\sqrt{4/N}$) which is the expected average number of particles if the other $N-1$ particles are distributed with uniformly over the sphere, but with a region of size $l=\sqrt{4/N}$ unoccupied.
  The inset shows $g(l)=\frac{2n'(l)}{Nl}$ where the derivative is with respect to $l$.
  This is identical to the two-point correlation function.
  It also shows the number variance $\sigma(l)$ as a function of $l$.
}\label{fig:sphere200}
\end{figure}

We first consider the Laughlin states again, this time on a sphere, rather than a plane, to avoid boundary effects.
This is a well established technique in fractional quantum Hall systems -- for details of the setup, we refer to the Methods section.
The results for the Laughlin states are presented in Figure~\ref{fig:sphere200} for $N=200$ particles.

On the sphere, we have chosen to measure the distance between particles in terms of the chord length $l=\left|\vec{r}_i-\vec{r}_j\right|$, between the particles $i$ and $j$,
where $\vec{r}$ is the 3D-position on the sphere
(this is the distance through the 3D ball, rather than over the surface of the sphere).
The chord length is the sphere analogue of the usual distance on the plane,
if one is interested in the surface area of the enclosed neighborhood,
since the surface area of a spherical cap of chord radius $l$ is precisely $A=\pi l^2$. We use units of length such that the sphere is a unit sphere and the maximal chord distance is $2$. 

In the upper panels of Figure~\ref{fig:sphere200}, we see both the $l$ and $n$ dependence of $p_n(l)$,
with the actual probabilities indicated by a color scale from $p=0$ (white) to $p=1$ (black).
Two support lines are drawn on top of $p_n(l)$:
The first (blue dashed line) is the average number of particles $n(l)=\sum_n np_n(l)$.
The second (green dashed line) is the particle density that one would expect if there was a sharp correlation hole around each particle of area $4\pi/N$,
outside of which the density is uniform. This turns out to be a very good approximation for low $q$ so that the support lines are actually on top of each other in the left and middle panels.

We now consider the full left panel ($q=1$) in more detail.
In the top left panel $a)$, we see that at any $l$, $n(l)$ and $\tilde n(l)$ are roughly identical.
This should be expected of a quantum fluid that is supposed to have no correlations beyond a characteristic length scale,
given by the fermi-exclusion hole of the electrons.
In the panel below $d)$, we show the two-point correlation function $g(l)$, normalized by the homogeneous density on the sphere.
This can be computed directly from the Monte Carlo generated data,
or alternatively as $g(l)=\frac{2n^\prime(l)}{Nl}$, which is what is shown.
The two-point correlator can be seen to flatten out after $l\approx0.3$, just beyond the correlation hole.

In the figure, we also show the variance $\sigma^2(l)=\sum_n n^2p_n(l)-n^2(l)$ of the counting statistics.
The variance is expected to fulfill $\sigma(l=0)=\sigma(l=2)=0$, since the number of particles is fixed. 
The lower the value of $\sigma(l)$, the more well defined is the number of neighbors within a the distance $l$.
In addition to $\sigma(l)$, a dashed support line shows the two-particle correlation function obtained from $\tilde{n}(l)$.

To demonstrate some of the features of the NCS more clearly, we look at the distributions of neighbors at some relevant fixed values of $l$. Here,
we choose three slices though the $p_n(l)$ plane, at $l=0.37$, $0.50$, $0.61$. 
These are the locations of significant dips and peaks in the two point functions at $q>1$, as can be seen in panels  $e)$ and $f)$. 
Plots of  $p_n(l)$ for these values of $l$ are shown in the lower panels. 
In these figures, $p_n(l)$ is plotted in blue.
While the binomial distribution $B_n$, which would be expected for $N=200$ randomly placed particles, is plotted in red for contrast.
We can easily see that even at $q=1$ (bottom left) the fluid is not randomly distributed,
even though it has homogeneous density. The particle number clearly has much lower variance than would be expected from $B_n$.
For $q=10$ (lower middle panel) the variance of $p_n$ is even more reduced as compared to $B_n$, as might be expected in an incompressible liquid with strong short range repulsive interactions.
It would be interesting to see if it is possible to obtain some analytical estimate of the expected behavior of $p_n$ as a function of $q$ in the liquid phase,
or even for generic liquids, but we have not managed this yet. 
The $l$ values of these slices are also indicated by a black bar in the middle panel $d)$,
and in the upper panel $a)$ the length of the slice is indicated by black dashed lines.
The standard deviation of the random (binomial) approximation is also indicated in the top panel as a red solid line.

\begin{figure}[t]
  \centering
  \includegraphics[width=1.00\linewidth]{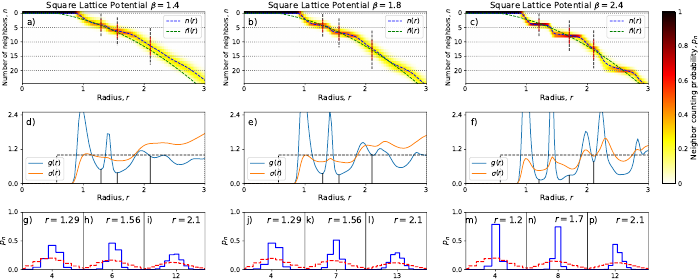}
\caption{Neighbor counting statistics for the modified Lennard-Jones potential, $N=100$, $\beta=14,11,24$.
   The blue line shows the average numbers of particles $n(r)$, within the radius $r$. 
  The green line is a guide to the eye, just as in Figure~\ref{fig:sphere200}.
  Here the precise form is $\tilde n(r)=\pi r^2-1$ (for $r>1/\sqrt{\pi}$) which is the expected average number of particles seen from the center of a square lattice, with unit lattice constant.
  The inset shows the two-point correlation function $g(r)$.}
\label{fig:square100}
\end{figure}

We now switch focus to the higher-$q$ states in the middle and right hand panels of Figure~\ref{fig:sphere200}. The setup of these panels is analogous to the one just described for the left panels. 
The $q=10$ state is a strongly correlated (but still liquid) quantum state,
whereas the $q=90$ state describes a Wigner crystal.
We find that the smooth $n(l)$ curve that was found for $q=1$ now has more features.
These features coincide with the peaks and troughs of both $g(l)$ and $\sigma(l)$,
and the larger the value of $q$, the more pronounced the peaks and troughs become.

We are now able to see the effects of the Wigner crystal at $q=90$, and of the onset thereof in the liquid at $q=10$.
For instance, there is spatial structure visible even in the density of the system for $q=90$ and clearly visible peaks and dips in the two point function for both $q=90$ and $q=10$.
In the $q=90$ case, in panel $f)$, we see that the Wigner Crystal has extended over the entire sphere, as the modulations in the two-point correlator $g(l)$ persist all the way to the maximal chord distance. We also note that the modulations in $g(l)$ are reflected by similar dips and peaks in the variance $\sigma(l)$. 

By considering $p_n(l)$ at a fixed $l$ we can directly visualize the shell structure of the hexagonal Wigner crystal.
We can for instance see in panel $m)$ that at $q=90$ there is roughly 75\% probability to find exactly $n=6$ neighbor particles within chord-radius $l=0.37$, 
and almost 100\% probability to find $n\in\{5,6,7\}$ neighbors within this chord-radius.
We interpret this as the Wigner crystal locally forming a hexagonal (triangular) lattice.
The sharp peak (in panel $m)$), together with the deep minima in $g(l)$,
tells us that there is a well defined region between the first and second shells.
Looking at higher values of $l$ we find distinct shell boundaries at $n\approx18$, 34, 54, $\ldots$ ($n=18$ is shown panel $p)$).
The number of particles that are expected to be found in the various shells, assuming a perfect hexagonal lattice structure, are $6$, 18, 36, 50, $\ldots$,
(see Figure~\ref{fig:LatticeFig}).  This is quite close to what we observe, but not identical at larger distances. However, first of all, it is mathematically impossible to place a perfect hexagonal crystal on a sphere, and secondly, even the shells in the perfect crystal become more closely spaced at longer distances, so they would be more difficult to distinguish there. 
We note once more that information about the \emph{average} number of particles that are in the various shells can in principle be extracted by integrating the two-point correlation function, but using neighbor counting statistics makes this much more transparent and also allows a direct view of the fluctuations and hence of the rigidity of the shells.

\subsubsection*{Modified Lennard-Jones potential}

For the modified Lennard-Jones potential we present data for $N=100$ particles in Figure~\ref{fig:square100}. The setup of this figure is analogous to that of Figure~\ref{fig:sphere200}.
We show plots for three values of the inverse temperature, $\beta=1.4$, $\beta=1.8$ and $\beta=2.4$.
The corresponding temperatures are all close to the transition temperature to the crystal state. In particular $\beta=1.4$ corresponds to a liquid system, which however already shows significant neighbor ordering, while $\beta=2.4$ is indistinguishable from a (just about) solid system in our fairly crude MC simulation. In our simulations, we have included a small harmonic trapping potential for the particles, in addition to the Lennard-Jones interaction. This adds a term $k\sum_{i=1}^{N}r_{i}^{2}$ to the energy, where we have taken $k=0.05$. This acts as an external pressure which helps the particles coalesce near the origin of the coordinate system. 

As this system is defined on a plane (just like the systems we considered when discussing the CD and CV), there are boundary effects in the density and higher order correlations. To make sure that we have the correct bulk averages, we choose to compute the NCS of the system by only counting the neighbors of the 25 particles closest to the system's center of mass.
We choose to only count one quarter of the particles, as this is the fraction that ensures that the particle that is furthest from the center (of the particles we consider), is approximately equally far from the center and the boundary.
Because of our selection of which particles to consider for the NCS we should not trust the NCS counting on radii that are larger than half of the radius of the particle cloud, or which contains more than one quarter of the particles.
This is also why panels $d)-f)$ only extend to $r=3$ and not over the full system size.

The left panel ($\beta=1.4$) has the highest temperature considered. The two point correlation function is normalized according to the (average) density of the square lattice, that is expected to form at low temperatures.
It is given by $g(r)=\frac{n^\prime(r)}{2\pi r}$.
Already here it is possible to see peaks in the two-point correlation function $d)$ at distances of $r=1$ and $r=1.44\approx\sqrt{2}$.
These two distances corresponds to the first two nearest neighbors on a square lattice.
By inspecting the NCS at radii $r=1.29$ and $r=1.56$ (subpanels $g)$ and $h)$), corresponding to the  minima in $g(r)$, we can investigate how well defined the shell structure of the square lattice is at this temperature.
Starting with $r=1.29$ we can see that there are on average 4 neighbors,
but the number may fluctuate between 3 and 5.
For the second shell, where we would expect 8 particles,
only 5-7 are found with reasonable probability.
We may then conclude that the lattice has only formed at the very shortest length scales.

Decreasing the temperature to $\beta=1.8$ (middle panel),
we see that, at the two minima of $g(r)$, in panels $j)$ and $k)$,
there are now 4-5 and 6-8 particles respectively.
Decreasing the temperature further to $\beta=2.4$ (right panel) clear steps can be seen at $n=4$ and $n=8$ in panel $c)$.
Indeed in the sub-panels $m)$ and $n)$ sharp peaks can be seen at $4$ and $8$ also in the distribution of the number of neighbor particles.

\begin{figure}
  \centering
  \includegraphics[width=.60\linewidth]{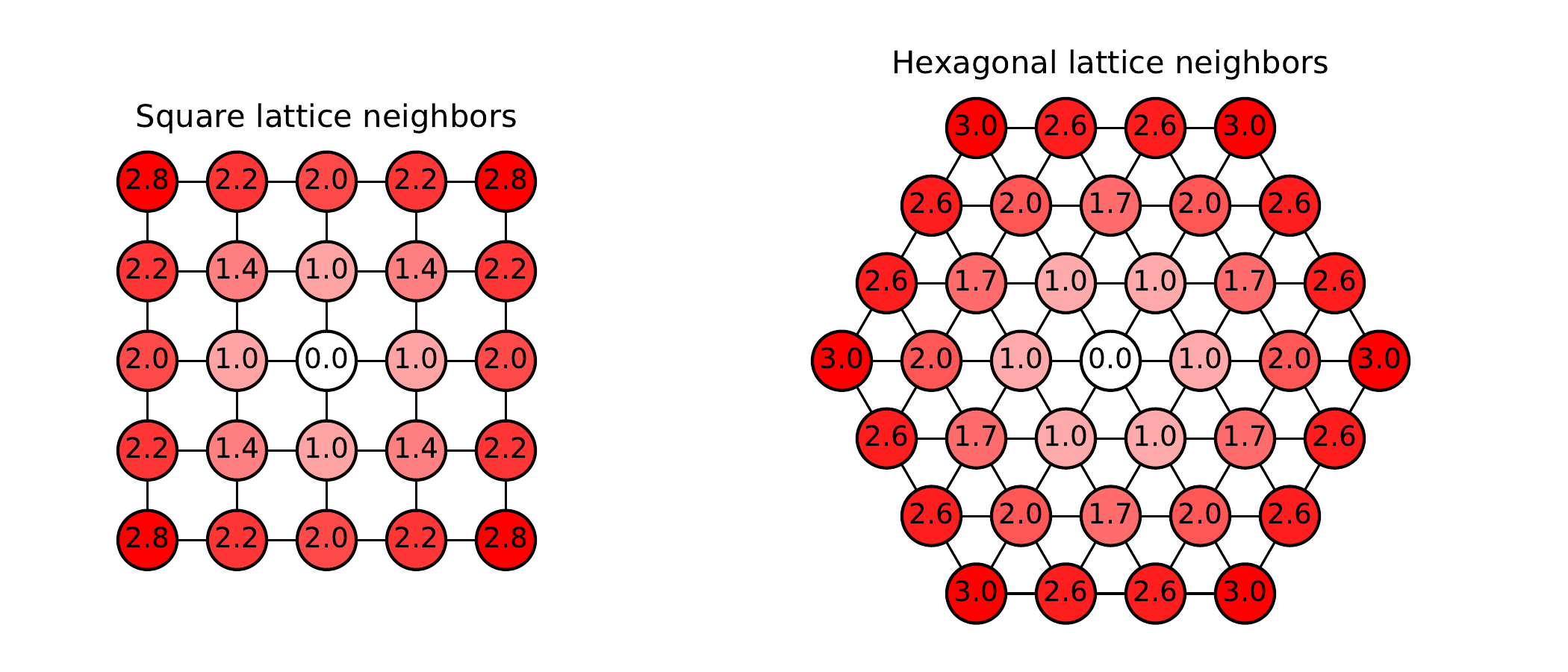}
  \caption{Distance to the neighbors on a square and triangular lattice.
    For the square the total neighbor counting is $4,8,12,20,24,\ldots$ at distances $1,\sqrt{2},2,\sqrt{5},2\sqrt{2},\ldots$.
    For the triangular lattice it is $6,12,18,30,36,\ldots$ at distances $1,\sqrt{3},2,\sqrt{7},3,\ldots$.
  }\label{fig:LatticeFig}
\end{figure}

Inspecting $g(r)$ for even larger values of $r$ we find that there is a minimum at roughly $r=2.1$.
Inspecting $p_n$ for this value of $r$ we find a distribution that is peaked at $n=12$ ( panels $i)$, $l)$ and $p)$ ) for the three values of $\beta$ considered. We can note that only the lowest temperature system ($\beta=2.4$) has a clear asymmmetry in the peak, showing that the disk of radius $r=2.1$ has only a small chance to contain fewer than $12$ particles.
In other words, it almost always fully contains the third shell of the square lattice.
Numbers higher than $12$ are considerably more likely, but this is not surprising as the next shell of the square lattice is quite close - the equilibrium position of its particles should be at $r\approx2.2$ (cf.\ Figure~\ref{fig:LatticeFig}).  

All in all, as exemplified by the two systems considered here, the NCS measure is a powerful complement to the more conventional two-particle correlation function.
Using the extra information from the NCS we can, not only, conclude that crystal patterns are (locally) forming, but we can also gain information about the rigidity or fluctuations of these crystals, as we get a direct view of the numbers of particles that are participating.

\section*{Discussion}

We started the paper with an argument against the physical relevance of Pauli crystals.
While the highest probability density configurations of a system of free Fermions may appear crystalline,
we showed that these have very low probability to be observed.
To demonstrate that the probability density of typical configurations that would be seen in an experiment is far from the maximal probability density, we constructed a histogram of the probability densities
from samples generated using MC importance sampling. It turns out that the functional form of the distribution,
which does not have a Gaussian shape (at finite system size),
can be fitted reasonably well by a one parameter curve
derived from a crude ad hoc model. 
It would be interesting to know to which extent this functional form 
will occur more generally in histograms of probability densities of physical systems.

We also studied the so called \emph{Configuration Density} (CD) method and found that it is highly biased towards the \emph{Seed Configuration} (SC) that is chosen for the method.
We demonstrated that a CD with crystalline features can appear where there are obviously no such features in the system and also that 
even physically unrealistic or bizarre SCs can be imprinted on CDs.
We argue that what is actually happening in the CD method is the following:
The process of rotating the sample configurations causes the density fluctuations, which are always present in these configurations,
to have a higher probability to be located near a seed position if they increase the density.
In effect a cluster of multiple particles can be located near the seed configuration in this way. 
Similarly, density fluctuations that decrease the density ("voids" in the configurations) will be preferentially located away from the seed positions.
As a result the average configuration density near the seed points is enhanced, which can give rise to the peaks in the CD,
even if the individual configurations contributing to the CD do not show any trace of a crystalline structure.
This was exemplified by considering the \emph{Configuration Variance} (CV) which measures the variance in the number of particles in a given region, around the SC, after the CD method has been applied.

We then introduced an alternative measure for the detection of spatial structure, \emph{Neighbor Counting Statistics} (NCS).
This is the full counting statistics of the neighborhood of a particle.
This measure incorporates information about the multi-particle correlations near a given particle in an intuitively understandable way.
Using the NCS we could demonstrate the onset of shell structures for particles interacting though a modified Lennard-Jones potential as well the formation of a Wigner Crystal in the $\nu=\frac{1}{90}$ Laughlin state.
The information gleaned from NCS was in accord with and complementary to the information we get more conventionally from observing peaks in the two-particle correlation function.
It would be of interest to test and apply this method further in the future,
for example for a more thorough examination of the many phases afforded\cite{Engel2007} by the modified Lennard-Jones potential.

We noted that the width of $p_n$ at a fixed radius (even for non-interacting fermions) was narrower than what would have been expected from randomly distributed points of the same density.
This in itself is not too surprising,
since the correlation hole due to the Pauli exclusion principle will penalize particle clustering.
Nevertheless, it would be of theoretical interest if bounds on $p_n(r)$ could be derived, for example
starting from knowledge of the two point correlation function $g(r)$.

\section*{Methods}

\subsection*{Laughlins's wave function on a sphere}
To avoid boundary effects, we modify the Laughlin state to be defined on a sphere with unit radius.
The Laughlin states can easily be adapted to the spherical geometry\cite{Haldane_1983} by applying a stereographic projection
\begin{equation*}
  z\to\tan\left(\frac\theta2\right)e^{i\phi},\quad\quad
  e^{-|z|^2/4\ell^{2}}\to \left[\cos\left(\frac{\theta}2\right)e^{-i\frac{\phi}2}\right]^{q(N-1)}.
\end{equation*}
The first transformation is applied to the polynomial coordinates in the Jastrow factor (i.e.~the polynomial factor) in \eqref{eq:Laughlin}, while the second is modifying the Gaussian part.
Here $\theta$ and $\phi$ are the azimuthal and polar coordinates of the sphere respectively.
After these transformations the Laughlin state reads
\begin{equation}
\Psi_{L}^{(q)}=\prod_{i<j}(u_iv_j-u_jv_i)^{q},\label{eq:Laughlin_spere}
\end{equation}
where $u=\sin\left(\frac{\theta}2\right)e^{i\frac{\phi}2}$ and $v=\cos\left(\frac{\theta}2\right)e^{-i\frac{\phi}2}$ are spinor variables.
We choose units of length such that the sphere has unit radius.

\subsection*{Deriving the self-describing distribution in Figure~\ref{fig:confs}}
For a probability density $\rho$, that depends on some (possibly multidimensional)
variable $x$ as $\rho(x)$ we require that
\begin{equation}
1=\int_{V_{x}}\rho(x)dx,\label{eq:Single-variables}
\end{equation}
where $V_{x}$ is the parameter-space of $x$.
Imagine that we only know $\rho(x)$ up to some scale factor (normalization) $\lambda$,
as $\rho(x)=\lambda f(x)$.
We now seek to construct the probability density $h(\rho)$ for the probability densities $\rho$ themselves.
We note that if $f(x)$ is one dimensional and monotonic,
then $d\rho=\lambda f^{\prime}(x)dx$.
By changing variables we may rewrite \eqref{eq:Single-variables} as 
\begin{equation}
1=\int_{0}^{\infty}\frac{\rho}{\lambda\left|f^{\prime}\left(f^{-1}\left(\frac{\rho}{\lambda}\right)\right)\right|}d\rho
=\int_{0}^{\infty}\frac{\rho}{\lambda}\gamma\left(\frac{\rho}{\lambda}\right)d\rho
=\lambda\int_0^{\infty}f\gamma(f)df
\propto\int_{0}^{\infty}h(\rho)d\rho.\label{eq:prob_norm}
\end{equation}
 In the last steps we introduced the shorthand notation $\gamma\left(\frac{\rho}{\lambda}\right)=\left|f^{\prime}\left(f^{-1}\left(\frac{\rho}{\lambda}\right)\right)\right|^{-1}$.
We note that a histogram over the distribution of $\rho$ is actually displaying $\rho\gamma(\rho)$, and not $\gamma(\rho)$ itself.
Thus $h(\rho)\propto \rho\gamma(\rho)$.
When expressed in logarithmic units $y=\ln \rho$, the integral is
\[
1=\int_{-\infty}^{\infty}h\left(e^{y}\right)\rho dy=
\int_{-\infty}^{\infty}h\left(e^{y}\right)e^{y}dy=
\int_{-\infty}^{\infty}\tilde{h}\left(y\right)dy.
\]
Here $\tilde{h}(y)=h(\rho)\rho$ is the observed density in logarithmic units.

We now set out to derive the form of $\tilde{h}(y)$ that should be expected in a system of many variables.
We choose the coordinates $x$ such that the maximum of the probability density is at $x=0$ (we assume that there is a unique maximum).
More drastically, we assume that the full probability density takes the form $\rho(x)=\lambda \exp(-\sum_{j=1}^{d}x_{j}^{2}/2\sigma^{2})$,
where $\lambda$ is an unknown normalization coefficient.
This is quite reasonable in a small neighborhood of the global minimum, but of course usually not beyond that small neighborhood.
However, if the probability density decreases quickly as soon as we move a particle away from the maximum density probability distribution,
it may be a reasonable approximation.
Note that we can choose the probability decay range $\sigma$ to be the same in all directions without loss of generality, 
since any rescaling of $x_{j}$ will only affect the normalization $\lambda$, which is unknown anyway.
We note that surfaces of constant $\rho$ will form $(d-1)$-spheres centered around $x=0$.
To map our problem back to the single variable case considered above, we can first integrate out these $d-1$ dimensions.
The probability integral is then 
\begin{align*}
1 & =\int_{V_{x}}\lambda e^{-\sum_{j=1}^{d}\frac{x_{j}^{2}}{2\sigma^{2}}}\prod _jdx_{j}=\lambda S_{d}\int_{0}^{\infty}e^{-\frac{r^{2}}{2\sigma^{2}}}r^{d-1}dr
\end{align*}
 where $S_{d}$ is the surface area of a $d$-dimensional unit sphere and $r^{2}=\sum_{i}x_{j}^{2}$.
 We stress already here that the peak of the distribution as a function of $r$ is not at $r=0$
 (where the most likely element is) but rather at $r=\sqrt{\sigma^{2}\left(d-1\right)}$
corresponding to $\rho=\lambda e^{-\frac{d-1}{2}}$.
Note also that the most likely value of $\rho$ is at first sight independent of $\sigma$,
however the $\sigma$ dependence is hidden in the normalization $\lambda$.
Next we change variables to $\rho$ using the transformations $d\rho=-\frac{r}{\sigma^{2}}e^{-\frac{r^{2}}{2\sigma^{2}}}dr=-\frac{r}{\sigma^{2}}\rho dr$,
giving 
\[
1=\lambda S_{d}\sigma^{d}2^{\frac{d}{2}-1}\int_{0}^{1}\left[-\ln(\rho)\right]^{\frac{d}{2}-1}d\rho.
\]
 We now would like to measure the distribution $h(\rho)\propto\left[-\ln\left(\frac1{\rho}\right)\right]^{\frac{d}{2}-1}$ in log-scale.
We thus change variables to $y=\ln \rho$ using $e^{y}dy=d\rho$ to obtain
\begin{align*}
1 & =\lambda S_{d}\sigma^{d}2^{\frac{d}{2}-1}\int_{-\infty}^{0}e^{y}\left(-y\right)^{\frac{d}{2}-1}dy
\end{align*}
such that we have  $\tilde h(y) =e^{y}(-y)^{\frac{d}{2}-1}$.
We may compute the averages of $y$ and $y^2$, giving 
\[
\left\langle y\right\rangle =-\frac{\Gamma\left(\frac{d}{2}+1\right)}{\Gamma\left(\frac{d}{2}\right)}=\frac{d}{2}~~~~~\mathrm{and}~~~
\left\langle y^{2}\right\rangle =-\frac{\Gamma\left(\frac{d}{2}+2\right)}{\Gamma\left(\frac{d}{2}\right)}=\frac{d}{2}\left(\frac{d}{2}+1\right).
\]
For this we use that $\int_{-\infty}^{0}\tilde h(y)dy =\Gamma\left(\frac{d}{2}\right)$,
$\int_{-\infty}^{0}y\tilde h(y)dy =-\Gamma\left(\frac{d}{2}+1\right)$ and $\int_{-\infty}^{0}y^2\tilde h(y)dy =\Gamma\left(\frac{d}{2}+2\right)$,
The variance of $y$ is then 
\[
\sigma_{y}^{2} = \left\langle y^{2}\right\rangle -\left\langle y\right\rangle ^{2}=\frac{d}{2}\left(\frac{d}{2}+1\right)-\left(\frac{d}{2}\right)^{2}=\frac{d}{2}.
\]
In Figure~\ref{fig:confs} we fit the function $\tilde h(y)$  to the probability densities obtained for non-interacting fermions in a harmonic trap, and for the Laughlin $\nu=1$ state,
by choosing $d$ such that the variances of the two distributions are equal.
This produces a surprisingly good fit.
On the other hand, the value of $d$ that best reproduces the observed density is not the physical dimension of the configuration space,
but this should probably have been expected, considering our rather simple minded model. 

\bibliography{confdens}

\section*{Acknowledgements}
This work was supported through SFI Principal Investigator Awards 12/IA/1697 and 16/IA/4524.
This work is part of the D-ITP consortium, a program of the Netherlands Organisation for Scientific Research (NWO) that is funded by the Dutch Ministry of Education, Culture and Science (OCW).
Special thanks to the Nordic Institute for theoretical Physics (NORDITA) and Centro De Ciencias De Benasque Pedro Pascual.
We are grateful for the use of the code package Hammer \cite{hammer}, which was used for the numerical simulations.

\section*{Author contributions statement}
J.K.S. and M.F. contributed equally to the main manuscript text and M.F. prepared the figures. Both authors reviewed the manuscript.

\section*{Additional information}
\subsection*{Competing interests}
The authors declare no competing interests.

\end{document}